\documentclass[%
superscriptaddress,
amsmath,
amssymb,
aps,
rmp,
floatfix,
]{revtex4-2}

\usepackage{geometry}
\usepackage{graphicx}
\usepackage{dcolumn}
\usepackage{bm}
\usepackage[mathlines]{lineno}

\usepackage{stmaryrd} 
\usepackage{placeins} 
\usepackage{subcaption} 
\usepackage{enumerate} 
\usepackage{ragged2e}
\usepackage{yhmath} 
\usepackage{enumitem}
\usepackage{amssymb}
\usepackage{amsthm}
\usepackage{stmaryrd}
\usepackage{diagbox}
\usepackage{array}
\usepackage{calrsfs}
\usepackage{colortbl}
\usepackage{makecell}
\usepackage{multirow}
\usepackage{geometry}
\usepackage{lipsum}
\usepackage{appendix} 
\usepackage{adjustbox}
\usepackage{pifont}
\usepackage[justification=raggedright]{caption}
\usepackage{parskip} 
\usepackage{todonotes}

\newcommand{\cmark}{\ding{51}}%
\newcommand{\xmark}{\ding{55}}%

\begin{document}

\preprint{APS/123-QED}

\title{Kendall Correlation Coefficients for Portfolio Optimization}

\author{Tomas Espana}
\email{tomas.espana@student-cs.fr}
\affiliation{Chair of Econophysics and Complex Systems, Ecole Polytechnique, 91128 Palaiseau Cedex, France}
\affiliation{LadHyX UMR CNRS 7646, \'Ecole Polytechnique, 91128 Palaiseau Cedex, France}

\author{Victor LeCoz}
\email{victor.lecoz@cfm.com}
\affiliation{Chair of Econophysics and Complex Systems, Ecole Polytechnique, 91128 Palaiseau Cedex, France}
\affiliation{LadHyX UMR CNRS 7646, \'Ecole Polytechnique, 91128 Palaiseau Cedex, France}
\affiliation{Quant AI Lab, 29 Rue de Choiseul, 75002 Paris, France}
\affiliation{Laboratoire de Math\'ematiques et Informatique pour la Complexit\'e et les Syst\`emes, CentraleSupélec, Universit\'e Paris-Saclay, 91192 Gif-sur-Yvette Cedex, France}
\affiliation{Capital Fund Management, 23 rue de l'Universit\'e, 75007 Paris, France}

\author{Matteo Smerlak}
\email{matteo.smerlak@cfm.com}
\affiliation{Capital Fund Management, 23 rue de l'Universit\'e, 75007 Paris, France}

\date{\today}

\begin{abstract}
Markowitz's optimal portfolio relies on the accurate estimation of correlations between asset returns, a difficult problem when the number of observations is not much larger than the number of assets. Using powerful results from random matrix theory, several schemes have been developed to `clean' the eigenvalues of empirical correlation matrices. By contrast, the (in practice equally important) problem of correctly estimating the eigenvectors of the correlation matrix has received comparatively little attention. Here we discuss a class of correlation estimators generalizing Kendall's rank correlation coefficient which improve the estimation of both eigenvalues and eigenvectors in data-poor regimes. Using both synthetic and real financial data, we show that these generalized correlation coefficients yield Markowitz portfolios with lower out-of-sample risk than those obtained with rotationally invariant estimators. Central to these results is a property shared by all Kendall-like estimators but not with classical correlation coefficients: zero eigenvalues only appear when the number of assets becomes proportional to the \emph{square} of the number of data points. 
\end{abstract}

\keywords{} 

\maketitle






\FloatBarrier
\section{Introduction}

Markowitz's modern portfolio theory is one of the most straightforward and widely used approaches to build diversified portfolios with a defined level of risk, taking as input the volatility and correlation of asset returns. A major challenge in applying this theory is the tendency to over-allocate low-variance modes (eigenvectors) of the correlation matrix. This can lead to catastrophic overinvestments in combinations of assets with low in-sample volatility but high out-of-sample risk \citep{bouchaud2009financial}. This issue arises from the inherent challenges of estimating correlation matrices in high-dimensional/data-poor settings. To achieve a reliable estimation of the $N(N-1)/2$ entries of the correlation matrix, a common rule of thumb suggests that the number of observations \( T \) should be at least 5 to 10 times the number of features \( N \). But in practice, the number of features (the size of the investment universe) often matches, or even exceeds, the number of observations. In this context, the estimation of the correlation matrix is intrinsically noisy. This considerably affects Markowitz's portfolios, since the small eigenvalues, which determine the least risky modes, are the ones most impacted by this measurement noise \citep{laloux1999noise}. Therefore, a strong motivation emerged to develop theoretical frameworks capable of extracting predictive signals from the data.

Initially encountered in mathematical statistics and later developed for its applications in nuclear physics, Random Matrix Theory (RMT) has become a powerful tool in quantitative finance. The pioneering work of \citet{laloux1999noise} revealed that the ''noise" in financial markets is pervasive but behaves according to the universal results of RMT. Extensive research has since been conducted to develop techniques for \textit{cleaning} (i.e.\ denoising) the eigenvalues of empirical covariance matrices; for comprehensive reviews, see \citet{bouchaud2009financial, Bun_2017}. These methods can be successfully applied to the construction of low-risk Markowitz portfolios \citep{bun2016my}. However, progress in this area has been constrained for two reasons: first, the best-performing cleaning methods are applicable only when the number of observations exceeds the number of features; and second, RMT results offer limited insights into how to clean eigenvectors.

The aim of this paper is to investigate an alternative approach, based on the notion of `generalized correlation coefficient', to cope with the problem of in-sample noise in eigenvalues and eigenvectors correlation matrices. We compare our methods to state-of-the-art cleaning schemes based on RMT, with very satisfactory results: the out-of-sample risk is lower for a wide class of investment strategies.

\section{Linear and generalized correlation coefficients}

\paragraph*{\textit{Linear Correlation.}}
The most common metric to measure the strength of the relationship between two random variables $X$ and $Y$ is Pearson's correlation coefficient $r$, also known as the \textit{linear} correlation. It is defined as the covariance between $X$ and $Y$, normalized by the product of their standard deviations: $r(X, Y) := \mathrm{Cov}(X, Y)/(\sigma(X) \sigma(Y))$. This moment-based correlation is popular for several reasons:

\begin{enumerate}
  \item It is straightforward to calculate: for many distributions, it is simple to calculate second moments (variances and covariances) and thus derive the correlation coefficient.
  \item Correlation and covariance are easy to manipulate under linear operations: the variance of any linear combination is fully determined by the pairwise covariances between the components. This property is of particular interest in portfolio theory.
  \item Correlation is a natural measure of dependence for multivariate normal distributions and, more generally, elliptical distributions. In these cases, knowledge of the marginal distributions and of the correlation matrix is sufficient to fully characterize the joint distribution. However, this is specific to elliptical distributions, an assumption which does not hold well for stock returns \citep{not_elliptical}.
\end{enumerate}

However, linear correlation also has shortcomings:

\begin{enumerate}
  \item It is not invariant under non-linear strictly increasing transformations $\psi$ of $X$ and $Y$, i.e., $r(\psi(X), \psi(Y)) \neq r(X, Y)$.
  \item The variances of $X$ and $Y$ must be finite. This causes problems when working with fat-tailed distributions.
  \item Since it depends on the marginals, the estimator of linear correlation is highly sensitive to outliers. In finance, this is a major issue since stock returns are known to be fat-tailed \citep{cont2001empirical}.
\end{enumerate}

\paragraph*{Rank Correlations}
These shortcomings motivate the introduction of correlation coefficients that are independent of the marginals $X$ and $Y$ and depend solely on the copula, that is, on the joint distribution of $(U, V)$ where $U = F_X(X)$ (resp. $V = F_Y(Y)$) with $F_X$ (resp. $F_Y$) the cumulative distribution function (cdf) of $X$ (resp. $Y$). Examples of such measures are Spearman's $\rho$ and Kendall's $\tau$, which are called \textit{rank} correlations. $\rho$ is simply defined as the linear correlation between $U$ and $V$, i.e. $\rho := r(F_X(X), F_Y(Y))$; $\tau$ can be expressed as a correlation of sign functions: $\tau(X, Y) := r(\mathrm{sign}(X - \widetilde{X}), \mathrm{sign}(Y - \widetilde{Y}))$, where $(\widetilde{X}, \widetilde{Y})$ is an independent copy of $(X, Y)$. The main known advantages of these rank correlations are their invariance under any strictly increasing transformations of $X$ and $Y$ and their ability to capture \textit{non-linear} monotonic dependencies (\citet{embrechts2002correlation}). However, since they are not moment-based, $\rho$ and $\tau$ do not support the same covariance manipulations as linear correlations.

To alleviate the issue of sensitivity, one can also calculate the linear correlation of clipped (i.e., winsorized) data: $r_c := r(X_c, Y_c)$, where the subscript notation $c$ denotes the clipped data. Practitioners frequently use this method to reduce the impact of outliers on the correlation matrix.

\paragraph*{Generalized Correlation Coefficients.}

Following \citet{daniels}, we call generalized correlation coefficient (GCC) a generalization of the Pearson correlation $r$ that measures a \textit{non-linear} dependence between two random variables $X$ and $Y$, defined as
\begin{align}
  \gamma^{\phi}(X,Y) = \frac{\mathbb{E}[\phi(X - \widetilde{X}) \phi(Y - \widetilde{Y})]}{\mathbb{E}[\phi(X - \widetilde{X})^2]^{1/2}\mathbb{E}[\phi(X - \widetilde{X})^2]^{1/2}}
\end{align}
where $(\widetilde{X}, \widetilde{Y})$ is an independent copy of $(X,Y)$ and $\phi$ is an odd function, called \textit{kernel}. If $(x_t)$ and $(y_s)$ are $T$ observations of $X$ and $Y$, the GCC can be estimated as
\begin{align}\label{eq:gcc}
  \widehat{\gamma}^{\phi}(X,Y) = \frac{\sum_{t < s} \phi(x_t - x_s) \phi(y_t - y_s)}{\left[\sum_{t < s} \phi(x_t - x_s)^2 \right]^{1/2}\, \left[\sum_{t < s}\phi(y_t - y_s)^2 \right]^{1/2}}.
\end{align}

When $\phi(x) = x$, Eq.\ (\ref{eq:gcc}) yields Pearson's usual estimator of linear correlation $\widehat{r}(X,Y)$. Similarly, when $\mbox{$\phi(x) = \text{sign}(x)$}$, Eq.\ (\ref{eq:gcc}) yields the standard estimator of Kendall's $\tau$, \newline $\mbox{$\widehat{\tau} = (\# \text{\{concordant pairs\}} - \# \text{\{discordant pairs\}}) \, / \, \# \text{\{pairs\}}$}$ where a pair of observations $(x_i, y_i)$ and $(x_j, y_j)$ is said to be \textit{concordant} if $\text{sign}(x_i - x_j) = \text{sign}(y_i - y_j)$ and \textit{discordant} otherwise. For $\rho$ and $r_c$, however, there are no functions $\phi$ such that their estimators are given by Eq.\ (\ref{eq:gcc}).

Note that Eq.\ (\ref{eq:gcc}) includes all pairs of observations $(x_i, x_j)$ and $(y_i, y_j)$ for $i < j$. For this reason, the numerator of Eq.\ (\ref{eq:gcc}) (i.e., the covariance) is a scalar product in a $\binom{T}{2}$-dimensional space, instead of a $T$-dimensional space as in the case of \textit{linear} correlation (this dimensionality reduction is due to the linearity). From this, one can readily prove that, in the multivariate case, the singularities of generalized correlation matrices (null eigenvalues) only appear when $N > T(T-1)/2 $, compared to $N > T$ for the sample correlation matrix. Thus, \textit{non-linear} generalized correlation matrices (GCMs) are particularly resistant to the appearance of zero eigenvalues in data-poor regimes. This property opens interesting perspectives for portfolio optimization when the ratio $q = N/T$ is close to or even greater than 1.

In the following, we will study the empirical properties of two \textit{non-linear} correlation coefficients: Kendall and the hyperbolic tangent kernel (i.e., $\phi = \tanh$). The latter is chosen because it behaves like the identity function near zero and the sign function for large values of $x$, thereby creating a `hybrid' GCC that bridges between Pearson and Kendall correlations. We will omit $\rho$ and $r_c$ since they are also \textit{linear} correlations and therefore share the same behavior than Pearson's $r$ when $q\to 1$. We summarize the main properties of the mentioned correlation coefficients in Table \ref{tab:correlation_methods}.

\begin{table}[htbp]
  \centering
  \captionsetup{position=top}
  \caption{Comparison of Correlation Coefficients}
  \definecolor{lightblue}{rgb}{0.9,0.95,1}
  \definecolor{green}{rgb}{0,0.6,0}
  \begin{adjustbox}{center,max width=\textwidth}
    \begin{tabular}{|>{\centering\arraybackslash}m{2.3cm}|>{\centering\arraybackslash}m{2.5cm}|>{\centering\arraybackslash}m{2.5cm}|>{\centering\arraybackslash}m{3.1cm}|>{\centering\arraybackslash}m{2.5cm}|>{\centering\arraybackslash}m{2.5cm}|}
      \hline
      \rule{0pt}{5ex}\raisebox{0.7ex}{Method}       & \raisebox{0.7ex}{\textbf{Pearson $r$}}              & \raisebox{0.7ex}{\textbf{Clip Pearson $r_c\,$}}                    & \raisebox{0.7ex}{\textbf{Spearman $\rho$}}                                                & \raisebox{0.8ex}{\makecell{\textbf{Kendall $\tau$}                                                                   \\ \fontsize{8}{12}\selectfont($\phi = \mathrm{sign}$)}}              & \raisebox{0.8ex}{\makecell{\textbf{GCC} \\ \fontsize{8}{12}\selectfont(general odd $\phi$)}}                           \\
      \hline
      \rule{0pt}{5.5ex}\raisebox{1.4ex}{Definition} & \raisebox{1.2ex}{${r}(X, Y)$}               & \raisebox{1.2ex}{${r}(X_c,  Y_c)$}             & \raisebox{1.2ex}{\fontsize{9}{12}\selectfont${r}\left(F_X(X), F_Y(Y)\right)$}     & \raisebox{1.5ex}{\fontsize{8}{12}\selectfont$
          \begin{array}{c}
            r(\mathrm{sign}(X - \widetilde X), \\
            \mathrm{sign}(Y - \widetilde Y))
          \end{array}$}
                                                    & \raisebox{1.5ex}{\fontsize{8}{12}\selectfont$
      \begin{array}{c}
            r(\phi(X - \widetilde X), \\
            \phi(Y - \widetilde Y))
          \end{array}$}                                                                                                                                                                                                                                                                                                                                                         \\
      \hline
      \rule{0pt}{5.5ex}\raisebox{1.4ex}{Estimator}  & \raisebox{1.2ex}{$\widehat{r}(X, Y)$}       & \raisebox{1.2ex}{$\widehat{r}(X_c, Y_c)$}      & \raisebox{1.2ex}{\fontsize{9}{12}\selectfont$\widehat{r}\left(R(X), R(Y)\right)$} & \raisebox{1.5ex}{\makecell{ $\widehat{\gamma}^{\phi} (X, Y)$ }}   & \raisebox{1.5ex}{\makecell{ $\widehat{\gamma}^{\phi} (X, Y)$ }}    \\
      \hline
      \rule{0pt}{2.5ex}Matrix Singularities         & \raisebox{1.5ex}{$N > T$}                               & \raisebox{1.5ex}{$N > T$}                                  & \raisebox{1.5ex}{$N > T$}                                                                     & \raisebox{1.5ex}{$N > \binom{T}{2}$}                      & \raisebox{1.5ex}{$N > \binom{T}{2}$}                       \\
      \hline
      \rule{0pt}{2.5ex}Independence to Marginals    & \raisebox{1ex}{\scalebox{1.2}{\textcolor{red}{\xmark}}} & \raisebox{1ex}{\scalebox{1.2}{\textcolor{orange}{\xmark}}} & \raisebox{1ex}{\scalebox{1.2}{\textcolor{green}{\cmark}}}                                     & \raisebox{1ex}{\scalebox{1.2}{\textcolor{green}{\cmark}}} & \raisebox{1ex}{\scalebox{1.2}{\textcolor{orange}{\xmark}}} \\
      \hline
    \end{tabular}
  \end{adjustbox}
  \parbox{\textwidth}{\justify \footnotesize \vspace{2ex} \textbf{Notes:} We use check marks (\cmark) when the independence to marginals is satisfied and cross marks (\xmark) otherwise (see \citet{embrechts2002correlation} for more details). The orange cross marks indicate that the method depends on the marginals but is not affected (or significantly less w.r.t.\ to red) by outliers/fat-tailed distributions. For Spearman's estimator, the notation $R$ denotes the rank operator, which assigns the rank to each observation within the sample.}
  \label{tab:correlation_methods}
\end{table}

\section{Cleaning schemes based on Random matrix theory} \label{sec:cleaning_schemes}

\paragraph*{Data Processing}

We consider a set of $N$ different financial stocks that we observe at a daily frequency, defining a vector of returns $\mathbf{x}_t = (x_{1t}, \ldots, x_{nt})$ at each day $ t = 1, \ldots, T$. Since volatilities of financial assets are known to be heteroskedastic \citep{bouchaud2003theory}, we therefore focus on the \textit{correlations} of stationarized data in order to study the systemic risk. To that end, we process the returns as in \citet{bun2016my}: (i) we remove the sample mean of each asset; (ii) we normalize each return by an estimate $\hat \sigma_{t}$ of its daily volatility: $\widetilde x_{it} := x_{it} / \hat \sigma_{t}$; for simplicity, we have chosen here the cross-sectional volatility $\hat \sigma_{t} = \sqrt{\sum_j x_{jt}^2}$ to remove a substantial amount of non-stationarity in the volatilities. The final $N \times T$ standardized return matrix is given by its entries $\mathbf{X}_{it} := \widetilde x_{it} / \sigma_i$ where $\sigma_i$ is the sample estimator of standard deviation of the $\widetilde x_i$. To a first approximation, these returns are now stationary.

\paragraph*{Eigenvalues of the Sample Correlation Matrix}

The most straightforward estimator of the 'true' correlation matrix, that we denote $\mathbf{C}$ henceforth, is the \textit{sample} estimator $\mathbf{E} := \frac{1}{T} \mathbf{X}\mathbf{X}^\top$. In what follows, we use the following eigen-decomposition: $\mathbf{E} = \sum_{k=1}^N \lambda_k \mathbf{u}_k \mathbf{u}_k^\top$ where $0 \leq \lambda_1 \leq \ldots \leq \lambda_N$ and $\mathbf{u}_1, \ldots, \mathbf{u}_N$ are respectively the eigenvalues and associated eigenvectors of $\mathbf{E}$. When $q = N/T \rightarrow 0$, i.e., when there is infinitely more data than stocks, one expects that $\mathbf{E} \rightarrow \mathbf{C}$. However, important distortions appear when $q = \mathcal{O}(1)$, even if $T \rightarrow \infty$. In fact, since the seminal work of \citet{marchenko1967distribution}, we know that the spectrum of $\mathbf{E}$ is a broadened version of that of $\mathbf{C}$, with an explicit $q$-dependent formula relating the two. In particular, as recalled in \citet{bun2016my}, small eigenvalues $\lambda_i$'s are underestimated, and big eigenvalues $\lambda_i$'s are overestimated; the greater the $q$ parameter, the more pronounced this distortion. Therefore, considering that Markowitz's theory takes the (inverse of the) sample covariance matrix as a direct input, it becomes clear why using the sample estimator $\mathbf{E}$ can lead to disastrous results. This underscores the importance of applying cleaning procedures to better approximate the true eigenvalues of \(\mathbf{C}\), before any application to portfolio construction.

\paragraph*{Two Eigenvalue Cleaning Schemes}

Among the many estimators proposed in the literature, we select two RMT-based eigenvalue cleaning schemes. The first method is relatively simplistic but effectively illustrates some interesting underlying RMT phenomena. The second method, more sophisticated, can be considered state of the art in this area. In the following, $\mathbf{\Xi}$ denotes a cleaned estimator of the true correlation matrix $\mathbf{C}$. Note that \(\mathbf{\Xi}\) and \(\mathbf{E}\) share the same eigenvectors; only the eigenvalues of \(\mathbf{E}\) are modified, and these modified eigenvalues will be denoted as \(\xi_k\).

\begin{enumerate}[leftmargin=10pt]
  \item \textbf{Eigenvalue Clipping $\mathbf{\Xi}^{clip}$} \citep{laloux2000random}:
        \vspace{0.2cm}
        \newline
        Keep the $N - \lceil N\alpha \rceil$ top eigenvalues and shrink the others to a constant $\zeta$ that preserves the trace, $\text{Tr}(\mathbf{\Xi}^{clip}) = \text{Tr}(\mathbf{E}) = N$:
        \vspace{-0.01cm}
        \begin{equation}
          \mathbf{\Xi}^{clip} := \sum_{k=1}^{N} \xi^{clip}_k \mathbf{u}_k\mathbf{u}_k^\top, \quad \xi^{clip}_{k} =
          \begin{cases}
             \zeta \,\,\,\,\,\, \text{if} \, k \leq \lceil N\alpha \rceil \\
            \lambda_k \,\,\, \text{otherwise}
          \end{cases}
        \end{equation}
        We assume that all empirical eigenvalues beyond the Marčenko-Pastur upper edge contain some signal and are therefore kept unchanged, while the remaining eigenvalues are considered to account for 'noise' and replaced by $\zeta$. This determines the choice of $\alpha$. Following standard practice, will refer to this method as \texttt{Clipped}\footnote{Note, however, that clipping in this sense is distinct from the process of data clipping, a.k.a.\ winsorization.}.
        One of the shortcomings of this cleaning methodology is that most of the cleaned eigenvalues are set to \(\zeta\), while the larger eigenvalues, which are known to be overestimated, remain unchanged. While this method uniformly cleans groups of eigenvalues (see the left column of Fig.\ \ref{fig:eigenvalues}), the following method adjusts each eigenvalue individually based on its specific value.

  \item \textbf{Rotationally Invariant Estimator $\mathbf{\Xi}^{rie}$} \citep{ledoit2011eigenvectors}:
        \begin{itemize}
          \item \underline{When $q < 1$:}
                \vspace{0.2cm}
                \newline
                The Marčenko-Pastur law describes how 'noise' affects the eigenvalues of \(\mathbf{E}\). The idea behind the Rotationally Invariant Estimator (RIE) is to apply the inverse transformation of this phenomenon to recover the true eigenvalues of \(\mathbf{C}\). This cleaning scheme is also known as \textit{optimal shrinkage} because it minimizes the mean squared error. The cleaned eigenvalues are given by:
                \vspace{-0.01cm}
                \begin{equation}
                  \xi^{\text{rie}}_k := \frac{\lambda_k}{|1 - q + q z_k s(z_k) |^2}
                \end{equation}
                where $q = N / T$, $s(z) := N^{-1} \text{Tr}(z\mathbf{I} - \mathbf{E})^{-1}$, and $z_k = \lambda_k - i\eta$. The parameter $\eta$ must be chosen to be small, but at the same time such that $N\eta \gg 1$. A good trade-off is to set $\eta = N^{-1/2}$ \citep{rie}. This method will be called \texttt{RIE}.
                \newline
                However, in cases where N is not extremely large (e.g., $N=500$), $\eta$ is not very small which induces a systematic downward bias in the estimator of small eigenvalues. Heuristically, the bias is corrected the following way \citep{bun2016my}:
                \begin{equation}
                  \xi_k^{\Gamma} := \xi^{\text{rie}}_k \times \text{max}(1, \Gamma_k) \geq \xi^{\text{rie}}_k
                \end{equation}
                For the explicit formula of $\Gamma_k$, refer to Box.\ 1 of \citet{bun2016my}. This regularized version will take the name \texttt{RIE $\Gamma$}. Note that, as expected, these cleaning schemes increase the small $\lambda_i$'s, decrease the big $\lambda_i$'s and leave the trace (almost) unchanged.

          \item \underline{When $q > 1$:} (adapted from \citet{Bun_2017}, Section 9.4)
                \vspace{0.2cm}
                \newline
                In this case, \(\mathbf{E}\) has \((N-T)\) zero eigenvalues. These singularities are not corrected by the previous RIE schemes and remain at zero. Therefore, applying RIE in this context primarily affects the higher eigenvalues, which are decreased and \(\text{Tr}(\mathbf{\Xi}^{\text{rie}})\) has (the most often) a downward bias. To address this, we propose the following simplistic method (\texttt{RIE + Id}):
                \vspace{-0.05cm}
                \begin{equation}\label{eq:rie_id}
                  {\mathbf{\Xi}}^{\text{id}} = \alpha \mathbf{I} + \mathbf{\Xi}^{\text{rie}}
                \end{equation}
                where $\alpha$ is chosen to recover the trace, $\text{Tr}({\mathbf{\Xi}}^{\text{id}}) = N$.
                Note that for \(q > 1\), due to singularities, there is limited literature on optimal cleaning schemes\footnote{\citet{Bun_2017} proposed to only rescale the $N-T$ null eigenvalues to a constant that preserves the trace. To easily associate the cleaned eigenvalues with their eigenvectors, we rescale all eigenvalues as in Eq.\ (\ref{eq:rie_id}).
                }. One particular issue is to detect exact zero modes of $\mathbf{C}$; if we follow this cleaning scheme, we systematically overestimate the volatility of these zero risk modes. In what follows, we will explore how GCMs, introduced earlier, can effectively tackle this issue.

        \end{itemize}


\end{enumerate}

\FloatBarrier
\section{Empirical Results on Synthetic Data} \label{sec:synthetic_data}

\paragraph*{Synthetic Data.}

For our simulations, we use daily data from 2000 to 2024 for 500 of the most liquid stocks from: America (USA, Canada), Europe, and Asia-Pacific (APAC). In this section, we focus only on American stocks (with similar results for the other regions). We generate the synthetic data as follows: (i) we estimate the \textit{correlation} matrix $\mathbf{C}$ from the daily returns of the 500 stocks; (ii) we generate synthetic returns from a multivariate Student distribution with mean zero, \textit{covariance} matrix $\mathbf{C}$ and three degrees of freedom; (iii) we apply the data processing steps described in the previous section to obtain the standardized returns $\mathbf{X}$. Even though this last step is not necessary in this case, we apply it to be consistent with the rest of the paper. We then estimate the sample correlation matrix $\mathbf{E}$ from $\mathbf{X}$ and conduct several studies on both eigenvalues and eigenvectors. In particular, we are interested on the behavior of GCMs when $q = \mathcal{O}(1)$.

\subsection{Eigenvalues}

This paragraph aims to provide a better understanding of the eigenvalues of the two GCMs previously mentioned: Kendall and the hyperbolic tangent kernel. The eigenvalues of $\mathbf{C}$ can be interpreted as the variances (i.e., the risk measure used in Markowitz's theory) of the portfolios corresponding to its associated eigenvectors. However, one cannot give a similar interpretation for the eigenvalues of GCMs. Therefore, it is interesting to compare the 'true' eigenvalues of $\mathbf{C}$ with those of GCMs. We also compare the eigenvalues of the latter with the estimated and cleaned eigenvalues of $\mathbf{E}$. We do this in two cases: $q = 0.5$ and $q = 2$.

\begin{figure}[ht]
  \centering
  \begin{minipage}{0.35\linewidth}
    \centering
    \includegraphics[width=\linewidth]{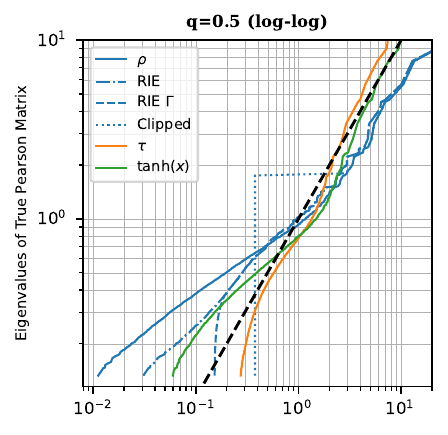}
  \end{minipage}
  \begin{minipage}{0.35\linewidth}
    \centering
    \includegraphics[width=\linewidth]{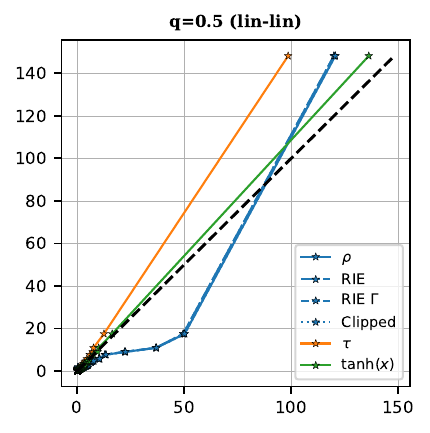}
  \end{minipage}
  \vspace{1em}
  \begin{minipage}{0.35\linewidth}
    \centering
    \includegraphics[width=\linewidth]{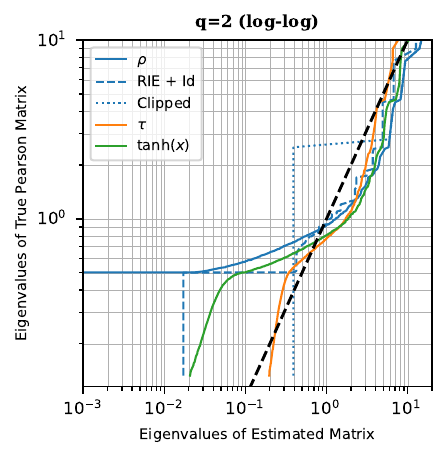} 
  \end{minipage}
  \begin{minipage}{0.35\linewidth}
    \centering
    \includegraphics[width=\linewidth]{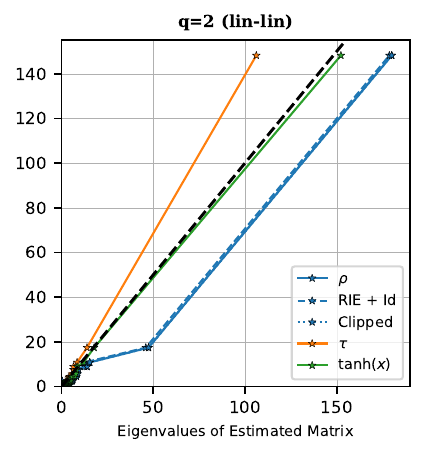} 
  \end{minipage}
  \caption{\justifying Eigenvalues of the true correlation matrix $\mathbf{C}$ ($y$-axis) against eigenvalues of the estimated correlation matrices ($x$-axis) for $q=0.5$ (first row) and $q=2$ (second row). The black dotted line corresponds to the identity function. The left column is in log-log scale and the right column in linear scale. The methods based on Pearson's correlation are in blue (the cleaning schemes are dotted). The eigenvalues of the estimated GCMs are in orange (Kendall) and green (hyperbolic tangent kernel). In the right column, we plot the data points (stars). In the left column, due to the high density of these points, we only plot the curve to enhance clarity.} 
  \label{fig:eigenvalues}
\end{figure}

In the first case, when $q=0.5$, there are twice as many observations as stocks. Therefore, we expect the eigenvalues of $\mathbf{E}$ to be reasonably close to the true ones, and those of the RIE and regularized RIE even closer. The upper left graph of Fig.\ \ref{fig:eigenvalues} confirms this expectation. Surprisingly, the eigenvalues of the GCMs are closer to the true ones than those of $\mathbf{E}$. 

As mentioned above, when $q \rightarrow 0$, we expect that $\mathbf{E} \rightarrow \mathbf{C}$, which implies that $\lambda_i$' s and $\xi_i$' s should align with the identity curve shown as a dotted black line in Fig. \ref{fig:eigenvalues}. However, this does not hold true for GCMs, as there is no reason for their eigenvalues to match those of the Pearson correlation matrix $\mathbf{C}$. In fact, for elliptical distributions, an analytical formula relates Pearson's and Kendall's correlation: $\rho = \sin(\frac{\pi}{2} \tau)$ \citep{lindskog2003kendall}. This relationship shows that $\tau$ is always lower than $\rho$ in absolute value. Therefore, we would expect the small Kendall eigenvalues to be larger than the corresponding small $\lambda_i$'s, and conversely, the large Kendall eigenvalues to be smaller. We indeed observe this in Fig.\ \ref{fig:eigenvalues}. By applying the aforementioned formula to each entry, it is therefore possible to transform a Kendall matrix into a Pearson correlation matrix, potentially providing a new method for estimating correlation matrices. However, this method was not selected due to the emergence of negative eigenvalues following the transformation. This phenomenon can be explained by the fact that the estimated Kendall correlation matrices do not always belong to the set of attainable Kendall correlation matrices for elliptical copulas \citep{MCNEIL2022105033}. 

In the second case, when $q = 2$, the situation is different as singularities appear: half of the sample eigenvalues $\lambda_i$'s are zero (we recall that the true eigenvalues cannot be zero, as there is no risk-free portfolio). These zero eigenvalues do not appear on the lower left graph of Fig.\ \ref{fig:eigenvalues}, which explains the plateau (in blue) that emerges from the left side. The smallest non-zero eigenvalue, $\lambda_{251}$, can be read on the $x$-axis just after the plateau ends, at approximately $0.03$.  Among the RMT cleaning schemes, the methods \texttt{RIE} and \texttt{RIE $\Gamma$} fail to correct for these zero eigenvalues, so we focus instead on the \texttt{RIE + Id} and \texttt{Clipped} methods. Once again, we observe that the eigenvalues of the GCMs are the closest to the true ones, especially for Kendall's method. This highlights the potential of GCMs for portfolio construction when $q > 1$. We recall that GCMs do not have singularities in this case since they appear under a \emph{quadratic scaling} of the number of observations, i.e. when $N > T(T-1)/2$. This means that with $N = 500$ stocks, the first singularity appears for only $T = 32$ observations.

Fig.\ \ref{fig:eigenvalues} shows similar results when we study the correlation matrix of stocks from Europe and APAC. The only slight difference is that the hyperbolic tangent kernel sometimes approximates the true eigenvalues more closely than Kendall. A trade-off between these two methods can be found by introducing the parametric function $\phi_{\beta}(x) = \tanh{(\beta x)}$ for $\beta > 0$. Notice that $\phi_{\beta} \underset{\beta \rightarrow \infty}{\longrightarrow} \text{sign}$, bridging the hyperbolic tangent and Kendall's kernel. In what follows we set $\beta = 1$, which is a reasonable choice when dealing with standardized data.

\FloatBarrier
\subsection{Eigenvectors}

While there is abundant literature on optimal cleaning of eigenvalues, the same cannot be said for eigenvectors. In this section, we continue our study on synthetic data, focusing on the comparison between the true and estimated eigenvectors of Pearson's correlation matrix and those of GCMs.

\paragraph*{Detecting Duplicate Stocks.}

\begin{figure}
  \centering
  \includegraphics[width=0.45\linewidth]{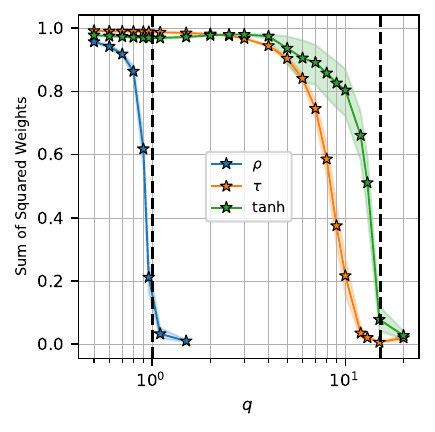}
  \caption{\justifying Sum of duplicate squared weights $\left(\mathbf{u}_1(1)\right)^2 + \left(\mathbf{u}_1(2)\right)^2$ ($y$-axis) against the sample size given by the ratio $q=N/T$ ($x$-axis), in semi-log scale, averaged over 100 realizations. The left vertical black dotted line represents $q = 1$ (indicating the appearance of singularities in $\mathbf{E}$) and the right one represents $q \approx 15$ (indicating the appearance of singularities in GCMs).}
  \label{fig:weights_doublons}
\end{figure}

We begin with a simple, albeit impractical, case that demonstrates the strong potential of GCMs to accurately estimate the eigenvectors of the true distribution. Among the 500 selected American stocks, we choose one that is traded in the USA and add the same stock traded on the Canadian market, creating a pair of nearly identical, or 'duplicate', stocks. These stocks are almost perfectly correlated, with a (Pearson) correlation greater than 0.98. Consequently, there exists an almost risk-free portfolio consisting of buying one stock and selling the other, albeit with no profit and loss. This portfolio is represented by the smallest eigenvector, $\mathbf{u}_1$. Assuming $\mathbf{u}_1(1)$ and $\mathbf{u}_1(2)$ are the weights associated with these two stocks and the eigenvectors are normalized, we expect $|\mathbf{u}_1(1)| = |\mathbf{u}_1(2)| = 1/\sqrt{2}$ and $\mathbf{u}_1(1) = -\mathbf{u}_1(2)$. We want to assess the robustness of Pearson's correlation and GCMs in detecting this pair of duplicate stocks as the sample size decreases. In Fig.\ \ref{fig:weights_doublons}, we plot $\left(\mathbf{u}_1(1)\right)^2 + \left(\mathbf{u}_1(2)\right)^2$ against the sample size. The remarkable performance of GCMs is evident as they can detect the duplicate stocks even with very few observations: with only 50 observations ($q = 10$), the sum of the squared weights exceeds 0.8 for the hyperbolic tangent kernel. In contrast, for Pearson's correlation, we observe that the curve drops much sooner, slightly before the frontier $q=1$, which marks the appearance of singularities in $\mathbf{E}$. Motivated by these results, we now focus on the ability of GCMs to accurately estimate all the eigenvectors of the true correlation matrix $\mathbf{C}$.

\paragraph*{Estimation of True Eigenvectors.}

In this paragraph, we set aside the duplicate stock mentioned above and return to the original 500 American stocks. In the perturbation theory of quantum mechanics, it is well-known that perturbed eigenvectors mix with the original ones. In fact, the closer the eigenvalues, the more the eigenvectors tend to mix. Therefore, one cannot compute the one-to-one overlap between the true and estimated eigenvectors and use it as a reliable metric of comparison. Instead, we use the fraction of common modes (henceforth referred to as FCM), a similarity metric between 0 and 1 that measures the volume of the common subspace spanned by \(n\) consecutive eigenvectors (see \citet{bouchaud2007large, benzaquen2017dissecting} for a more detailed description). Let us introduce a few additional notations. Let \(\mathbf{v}_1, \ldots, \mathbf{v}_N\) be the eigenvectors of \(\mathbf{C}\), where \(\mathbf{v}_N\) corresponds to the largest eigenvector. Let \(\mathbf{P}\) be the \(N \times N\) matrix whose columns are \(\mathbf{v}_N, \ldots, \mathbf{v}_1\), and \(\widehat{\mathbf{P}}\) the \(N \times N\) matrix whose columns are \(\mathbf{u}_N, \ldots, \mathbf{u}_1\). The FCM is constructed as follows: (i) compute the overlap matrix \(\mathbf{U} = \mathbf{P}^\top \widehat{\mathbf{P}}\); (ii) crop the matrix \(\mathbf{U}\) to keep only the first \(n\) rows and columns, where \(n \in \llbracket 1, N \rrbracket\); (iii) compute its singular values \(\{ w_k \}_{1 \leq k \leq n}\); (iv) compute their geometric mean \(\left( \prod_{k=1}^n w_k \right)^{1/n}\) (the so-called fraction of common modes) and plot it as a function of \(n\).

\begin{figure}
  \centering
  \begin{subfigure}{0.45\linewidth}
    \centering
    \includegraphics[width=\linewidth]{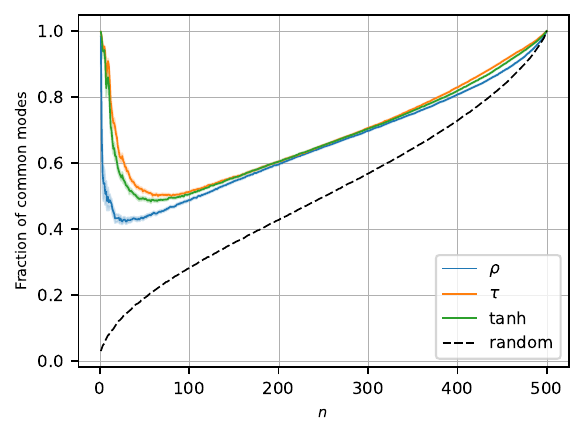}
    \caption{$n$ first eigenvectors (large modes)}
    \label{fig:focm_large}
  \end{subfigure}
  \begin{subfigure}{0.45\linewidth}
    \centering
    \includegraphics[width=\linewidth]{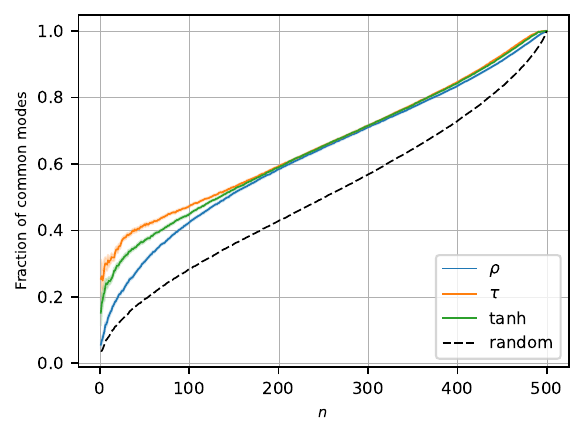}
    \caption{$n$ last eigenvectors (small modes)}
    \label{fig:focm_small}
  \end{subfigure}
  \caption{Fraction of common modes averaged over 50 realizations for $q=1$.}
  \label{fig:focm}
\end{figure}

In Fig.\ \ref{fig:focm}, we plot the FCM for the first and last $n$ eigenvectors to assess the quality of both large and small estimated eigenvectors, for sample sizes such that $q = 1$. The results are compared with a random strategy, which consists of randomly picking, one after the other, the elements of an orthogonal basis of $\mathbb{R}^N$ and computing the FCM. Thus, the random strategy serves as a benchmark: curves above it indicate that at least some of the true information is captured in the estimates. In both graphs, we observe that GCMs perform remarkably better and are able to capture the true eigenvectors more accurately than Pearson's correlation, especially Kendall. Note that for the large eigenvectors (Fig.\ \ref{fig:focm_large}), the FCM curve starts close to 1. This is because the market mode, which has the largest eigenvalue (see the right column of Fig.\ \ref{fig:eigenvalues}), is easily captured by all methods. However, Pearson's FCM drops sharply afterward, whereas the GCMs maintain a high FCM on the next 10 eigenvectors. In both cases, these results suggest that GCMs are excellent candidates for reliably estimating the eigenvectors of the true correlation matrix $\mathbf{C}$ (see Appendix \ref{app:focm} for a more detailed study).

\paragraph*{Estimation of Small Modes.}

\begin{figure}[ht]
  \centering
  \begin{subfigure}{0.45\linewidth}
    \centering
    \includegraphics[width=\linewidth]{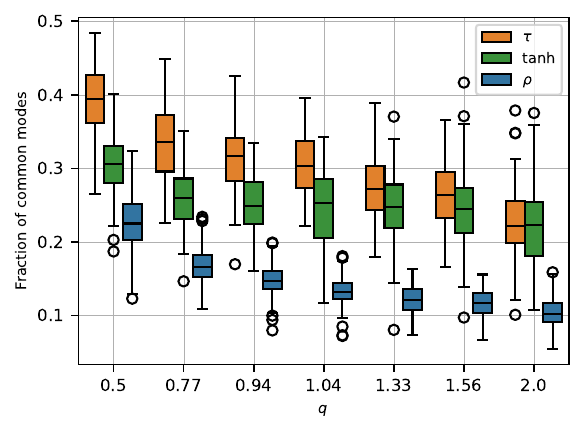}
    \caption{Correlation Matrix}
    \label{}
  \end{subfigure}
  \begin{subfigure}{0.45\linewidth}
    \centering
    \includegraphics[width=\linewidth]{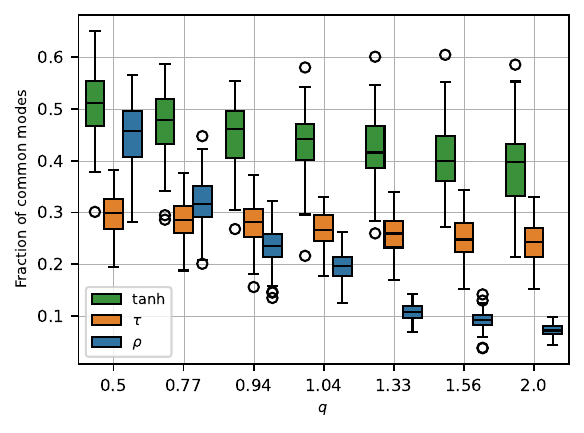}
    \caption{Covariance Matrix}
    \label{fig:boxplot_focm_cov}
  \end{subfigure}
  \caption{\justifying Fraction of common modes over the 10 smallest eigenvectors ($y$-axis) against the sample size given by the ratio $q=N/T$ ($x$-axis) for 100 realizations.}
  \label{fig:boxplot_focm}
\end{figure}

Markowitz’s theory, also known as the mean-variance model, seeks an optimal portfolio that minimizes variance (i.e., risk) while satisfying a minimum expected return constraint. Due to this focus on minimizing variance, Markowitz’s theory assigns a large weight to the eigenvectors of the \textit{covariance} matrix with the smallest eigenvalues. Therefore, accurately estimating the smallest modes of the covariance matrix is crucial for constructing a portfolio with minimal risk; we extend our previous study to small modes only. Fig.\ \ref{fig:boxplot_focm} displays the fraction of common modes over the 10 smallest modes for both the \textit{correlation} and the \textit{covariance} matrix, as the sample size decreases. For the correlation matrices, the results are similar to those of Fig.\ \ref{fig:focm_small}, with GCMs, especially Kendall, outperforming Pearson across a wide range of sample sizes. However, the normalization that converts the correlation matrix into a covariance matrix (using the true standard deviations of the distribution) disrupts the eigenvectors of the GCMs, as shown in Fig.\ \ref{fig:boxplot_focm_cov}. This disruption is particularly notable for Kendall whose performance, for $q<1$, falls below Pearson\footnote{Similar results are observed for European stocks. For APAC stocks, GCMs consistently perform better for both correlation and covariance.}. This loss of information is probably explained by the fact that GCMs are not moment-based (i.e., there is a bias between the Pearson and the GCM eigenvalues), and therefore the normalization alters the eigenvectors. A solution to this problem is discussed in the following section.

\FloatBarrier
\section{Optimal shrinkage, GCMs and cross-validation for portfolio optimization}

\subsection{Isotonic Cross-Validation Covariance}

The previous study has shown that GCMs are better than Pearson's correlation for estimating the eigenvectors of the true correlation matrix. However, because of differences in the eigenvalues, the quality of the covariance estimate is diminished. In this paragraph, we propose a method that combines the best of both worlds: leveraging the strength of GCMs for eigenvectors and the moment-based interpretability of Pearson eigenvalues. This method is largely inspired by the Isotonic Cross-Validation Covariance (ICVC) introduced by \citet{bartz2016cross}. The ICVC is a method that uses cross-validation to shrink the covariance matrix in a non-linear way. It has proven to outperform other shrinkage methods and also factor analysis-based methods in terms of out-of-sample portfolio risk \citep{bartz2016cross}.

More specifically, keeping the same notations as in Section \ref{sec:cleaning_schemes}, the ICVC method involves the following steps. The returns given by the \(N \times T\) matrix \(\mathbf{X}\) are divided into 10 folds. For each of the \(K \in \llbracket 1, 10 \rrbracket\) folds: (i) on the validation set, we compute the eigenvectors \(\mathbf{a}_1^K, \ldots, \mathbf{a}_N^K\) of the sample covariance matrix; (ii) we compute the variance of these eigenvectors on the test set, yielding $\mu_1^K, \ldots, \mu_N^K$. We then average these variances over the folds, \(\mu_i = \frac{1}{10}\sum_{K=1}^{10} \mu_i^K\), and perform an isotonic regression, which yields the new eigenvalues \(\mu_i^{\text{iso.}}\). The eigenvectors of the shrunk matrix are those of the empirical correlation matrix \(\mathbf{E}\). Finally, \(\mathbf{\Xi}^{\text{ICVC}} = \sum_{i=1}^N \mu_i^{\text{iso.}} \mathbf{u}_i \mathbf{u}_i^\top\).

We propose an adaptation of this method to GCMs: instead of selecting the eigenvectors of the sample covariance matrices, we choose those of the estimated GCMs. This modifies the eigenvectors \(\mathbf{a}_i^K\) in step (i) and the final eigenvectors \(\mathbf{u}_i\). Despite this change, the eigenvalues remain moment-based since they are computed as the variances of the eigenvectors on the test set. Therefore, we manage to leverage the best of both worlds. For Kendall, this method will be referred to as \texttt{Kendall ICVC}.

\subsection{Portfolio Optimization}

This section is largely inspired by the work of \citet{bun2016my}, with the aim of replicating and extending their study on the out-of-sample risk of Markowitz portfolios.
\vspace{-0.5cm}

\paragraph*{Markowitz Portfolio Theory.}

Markowitz portfolio theory is based on a mean-variance optimization to determine an asset allocation that minimizes the overall quadratic risk of the portfolio while ensuring a minimum expected target return. This problem can be formulated as a quadratic optimization problem with a linear constraint. More precisely, we seek the portfolio \( \mathbf{w}^* \) that solves:
\begin{equation}
  \min_{\mathbf{w} \in \mathbb{R}^N} \frac{1}{2} \mathbf{w}^\top \mathbf{\Sigma} \, \mathbf{w} \quad \mathrm{s.t. \,\,} \mathbf{w}^\top \mathbf{g} \geq \mathcal{G}
  \label{eq:opti_problem}
\end{equation}
where $\mathbf{\Sigma}$ is the estimated covariance matrix, $\mathbf{g}$ a $N$-dimensional vector of predictors (assumed to be deterministic) and $\mathcal{G}$ the expected gain (we set $\mathcal{G} = 1$ henceforth). The optimal portfolio can easily be found introducing a Lagrange multiplier and leads to a linear problem where the matrix $\mathbf{\Sigma}$ has to be inverted. In particular, the composition of the least risky portfolio has a large weight on the eigenvectors of $\mathbf{\Sigma}$ with the smallest eigenvalues. The solution is given by
\vspace{-0.01cm}
\begin{equation}
  \mathbf{w}^*= \frac{\mathbf{\Sigma}^{-1} \mathbf{g} }{\mathbf{g} ^\top\mathbf{\Sigma}^{-1} \mathbf{g}}
\end{equation}

In our case, $\mathbf{\Sigma}$ will either be the cleaned covariance matrix $\mathbf{\Sigma}_{ij} = \sigma_i \sigma_j \mathbf{\Xi}_{ij}$ for $(i,j) \in \llbracket 1, N \rrbracket$ or the GCM covariance matrix. Note that, depending on the application, different and additional constraints can be incorporated into the optimization problem (\ref{eq:opti_problem}). Commonly, we can impose that the sum of all portfolio weights equals one (no available cash) or restrict the weights to be non-negative (no short-selling). However, these constraints limit the admissible set and significantly reduce the influence of the covariance estimate. To fully leverage the quality of the estimation methods, we therefore apply only the constraint of a minimal expected gain.

\paragraph*{Stocks and Data Processing.}

We follow the same processing scheme as outlined in Section \ref{sec:cleaning_schemes} and use the same 500 American stocks mentioned in Section \ref{sec:synthetic_data}. We choose $T = 1000$ (4 years) for the training (in-sample) period, i.e., $q = 0.5$ and $T_{\mathrm{out}} = 60$ (3 months) for the out-of-sample test period. More precisely, we split the total length of our time series $T_{\mathrm{tot}} \approx 6300$ into $n$ consecutive, non-overlapping samples of length $T_{\mathrm{out}}$. Since the in-sample has length $T$, the integer $n$ is given by $n := \left\lfloor \frac{T_{\mathrm{tot}} - T - 1}{T_{\mathrm{out}}} \right\rfloor.$

\paragraph*{Four Trading Strategies.}

In order to assess the robustness of our results in different market situations, we consider the following four trading strategies:
\begin{enumerate}[leftmargin=10pt]
  \item \textbf{Minimum Variance Portfolio:} $\mathbf{g} = (1,...,1)^\top$.
  \item \textbf{Omniscient Portfolio:} The exact realized return on the next out-of-sample period is known. The portfolio is given by $g_i = \mathcal{N} \widetilde x_{it}\left(T_{out}\right)$ where $x_{it}(\tau) = (P_{i,t+\tau} - P_{i,t}) / P_{i,t}$ with $P_{i,t}$ the price of the \textit{i}th asset at time \textit{t} and $\widetilde x_{it} = x_{it}/\hat \sigma_{t}$.
  \item \textbf{Mean-Reversion Portfolio:} We mean-revert the return of previous day (daily rebalancing), \newline $g_i = -\mathcal{N}\widetilde x_{it}$.
  \item \textbf{Random Long-Short Portfolio:} $\mathbf{g}=\mathcal{N} \mathbf{v}$, where the coefficients of $\mathbf{v}$ are independent standardized Gaussian random variables such that $\|\mathbf{v}\|=1$.
\end{enumerate}

The normalisation factor $\mathcal{N}$ is chosen such that $\|\mathbf{g}\| = \sqrt{N}$ to ensure $w_i \sim \mathcal{O}\left(N^{-1}\right)$ for all $i$, thus it may vary between methods\footnote{In \citet{bun2016my}, the authors set $\mathcal{N} = \sqrt{N}$ across all methods, which, with their data, is sufficient to ensure $w_i \sim \mathcal{O}(N^{-1})$.}.

\paragraph*{Out-of-Sample Risk.}

The out-of-sample risk $\mathcal{R}^2_{\text{out}}$ is obtained by computing the quadratic risk of the portfolio $\mathbf{w}$ on the $n$ consecutive out-of-sample periods by
\begin{align}
  \mathcal{R_{\text{out}}}^2(t, \mathbf{w}) := \frac{1}{T_{\mathrm{out}}} \sum_{\tau = t+1}^{t + T_{\mathrm{out}}} \left( \sum_{i=1}^N \mathbf{w}_i \widetilde x_{i \tau} \right)^2
\end{align}

\begin{figure}[ht]
  \centering
  \includegraphics[width=0.65\linewidth]{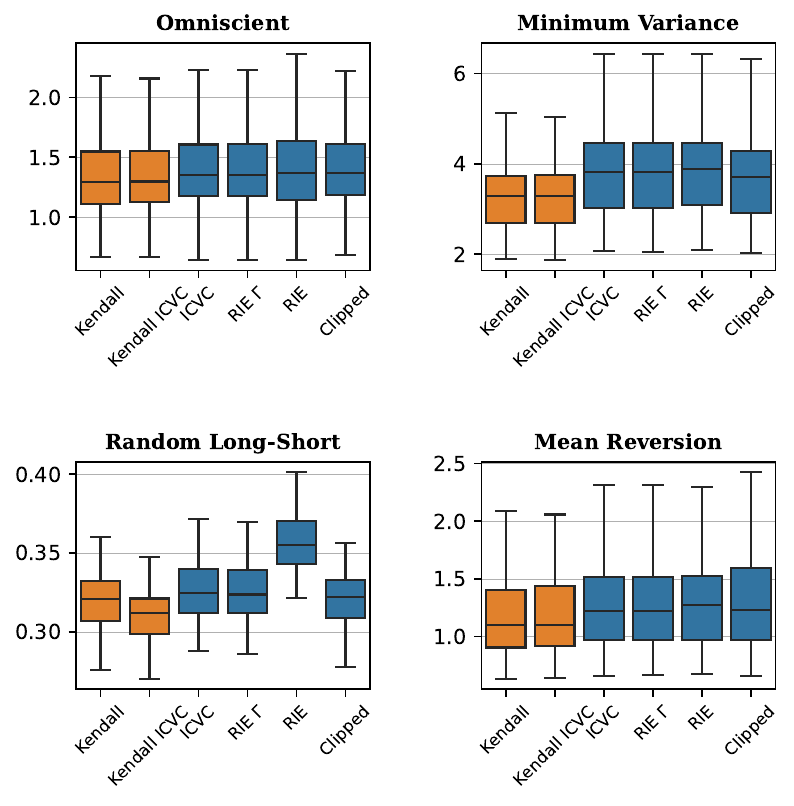}
  \caption{Annualized volatility $\left \langle \sigma_{\mathrm{out}} \right \rangle$ (in \%) of the different strategies for $q=0.5$ (American stocks).}
  \label{fig:ofs_risk}
\end{figure}

In Fig.\ \ref{fig:ofs_risk}, we present the boxplots of the annualized volatilities associated with the \( n \) consecutive risks \(\mathcal{R}^2_{\text{out}}\). The results for the cleaning schemes based on RMT (i.e., \texttt{RIE}, \texttt{RIE $\Gamma$}, and \texttt{Clipped}) are consistent with those reported by \citet{bun2016my}. We observe that the two Kendall-related methods consistently demonstrate lower out-of-sample risk across all four trading strategies. The results for the hyperbolic tangent kernel were omitted as they were indistinguishable from those of Kendall. This study coroborates the previous empirical findings suggesting that Kendall is an excellent candidate for portfolio construction. In particular, \texttt{Kendall ICVC} stands out as the best candidate among all the methods. This conclusion is further supported by a more exhaustive study in Appendix \ref{app:tables_risk}, where we vary the sample size \( q \in \{0.5, 1, 2\} \) for America, Europe, and APAC.

\FloatBarrier
\section{Conclusion}

We have introduced generalized correlation coefficients defined by an odd kernel function. Specific kernels yield well-known coefficients such as Pearson's and Kendall's correlation coefficients. In a multivariate setting, we have demonstrated that generalized correlation matrices are more robust to the appearance of singularities than Pearson's correlation matrix: null eigenvalues appear when $N = \mathcal{O}(T^2)$ instead of $N = \mathcal{O}(T)$. We have leveraged this property to reliably estimate correlation matrices in high-dimensional/data-poor settings. Specifically, we have focused on Kendall's correlation coefficient and the hyperbolic tangent kernel. Our study on synthetic multivariate Student data showed that, when the number of observations is low, the eigenvalues of GCMs, although biased in theory, are closer to the true Pearson eigenvalues than those obtained from standard Pearson coefficients. We have shown that these GCMs are also more accurate in estimating the eigenvectors of the true correlation matrix than the usual Pearson coefficients. 

The choice of an estimator often boils down to a trade-off between bias and variance. When estimating the spectrum of covariance matrices, different choices of estimators can lead to more accurate estimations of either small or large eigenvalues. Thus, depending on the problem considered, different choices of GCMs should be made. In finance, if the objective is to minimize risks, estimating a covariance matrix that is accurate on the smallest modes is crucial. We find that a minimum-risk portfolio estimated using Kendall's correlation matrix outperforms the state-of-the-art cleaning schemes based on RMT out of sample. By construction, the Kendall estimator assigns the same weight to small and large co-movements of a pair of assets, whereas the usual Pearson coefficient over-weighs large moves, making it blind to small modes when there are few observations. In other words, the bias in the risk measure resulting from estimating correlations on a distorted representation of the data is overcompensated by the enhanced robustness of the estimator in data-poor regimes. Finally, we show that the use of this Kendall estimator combined with the isotonic cross-validation covariance method outperforms all alternative approaches.

Future research could focus on the theoretical properties of GCMs, such as the asymptotic behavior of the eigenvalues and eigenvectors in simple models like the spiked covariance model, or on developing new cleaning schemes based on GCMs.

\begin{acknowledgments}
We would like to express our gratitude to Jean-Philippe Bouchaud, Konstantin Tikhonov and Pierre Bousseyroux, who contributed to our research through fruitful discussions. We are also grateful to Joël Bun for his valuable insights and for verifying our data processing methodology to ensure consistency with his own approach, as described in \citet{bun2016my}.
\newline
This research was conducted within the Econophysics \& Complex Systems Research Chair, under the aegis of the Fondation du Risque, the Fondation de l'Ecole Polytechnique, the Ecole Polytechnique and Capital Fund Management.
\end{acknowledgments}


\bibliographystyle{apsrev4-2} 
\bibliography{sample}


\appendix

\newpage
\FloatBarrier
\section{Kendall Eigenvalue Clipping}

We introduce an additional cleaning scheme based on RMT, which is an equivalent of the eigenvalue clipping scheme \(\mathbf{\Xi}^{\text{clip}}\) for Kendall's matrix. The eigenvalues of Kendall's matrix are known to follow an affine transformation of the Marčenko-Pastur distribution \citep{bandeira2017marvcenkopastur}. Therefore, we can clip the eigenvalues of Kendall's matrix: those above the theoretical upper edge remain unchanged, while the others are replaced by a constant value that preserves the trace. We include this method, which we call \texttt{Kendall Clipped}, in the tables of Appendix \ref{app:tables_risk}. Notably, it is interesting to compare \texttt{Clipped} and \texttt{Kendall Clipped}. Our results suggest that the latter outperforms the former.

\FloatBarrier
\section{Fraction of Common Modes}\label{app:focm}
\begin{figure}[ht]
  \centering
  \begin{minipage}{0.35\linewidth}
    \centering
    \includegraphics[width=\linewidth]{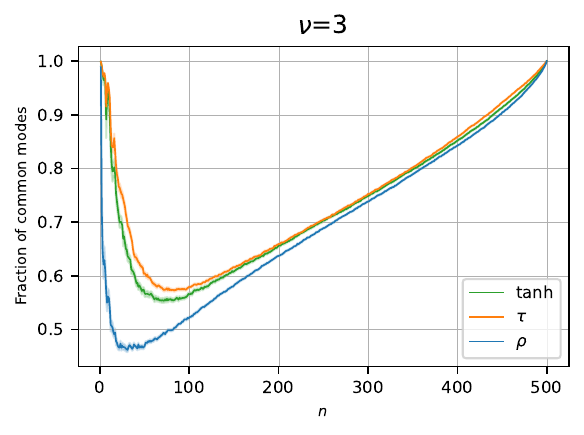}
  \end{minipage}
  \begin{minipage}{0.35\linewidth}
    \centering
    \includegraphics[width=\linewidth]{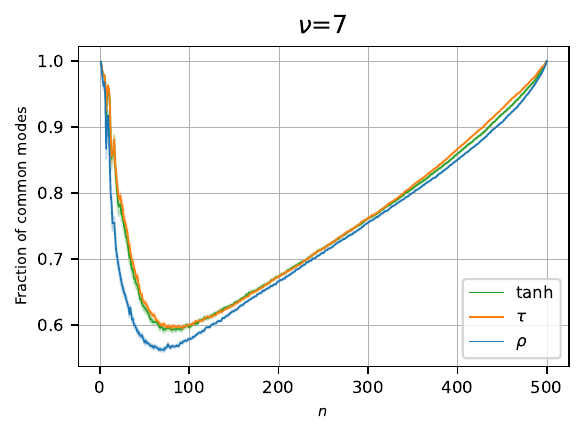}
  \end{minipage}
  \vspace{1em}
  \begin{minipage}{0.35\linewidth}
    \centering
    \includegraphics[width=\linewidth]{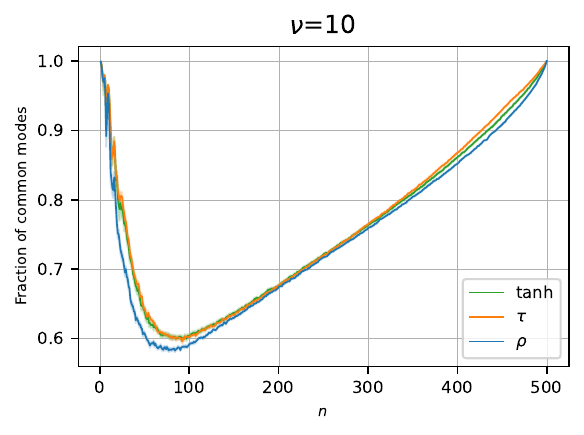} 
  \end{minipage}
  \begin{minipage}{0.35\linewidth}
    \centering
    \includegraphics[width=\linewidth]{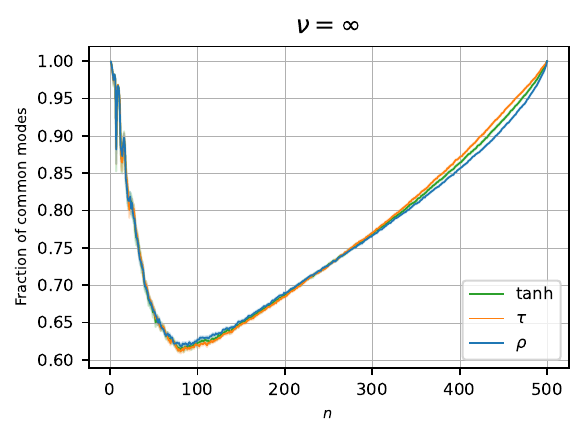} 
  \end{minipage}
  \caption{ Fraction of common modes averaged over $50$ realizations for $q = 0.5$.} 
  \label{fig:focm_0.5}
\end{figure}

\begin{figure}[ht]
  \centering
  \begin{minipage}{0.35\linewidth}
    \centering
    \includegraphics[width=\linewidth]{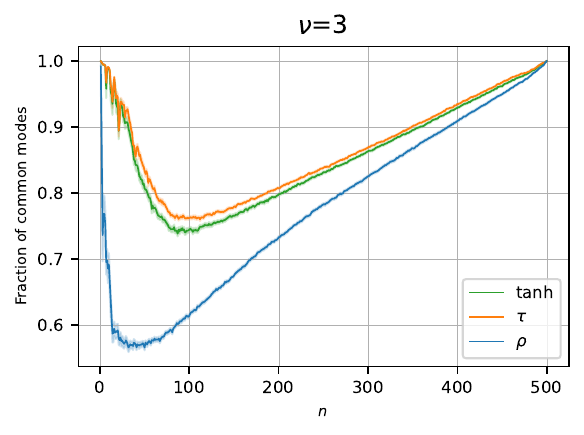}
  \end{minipage}
  \begin{minipage}{0.35\linewidth}
    \centering
    \includegraphics[width=\linewidth]{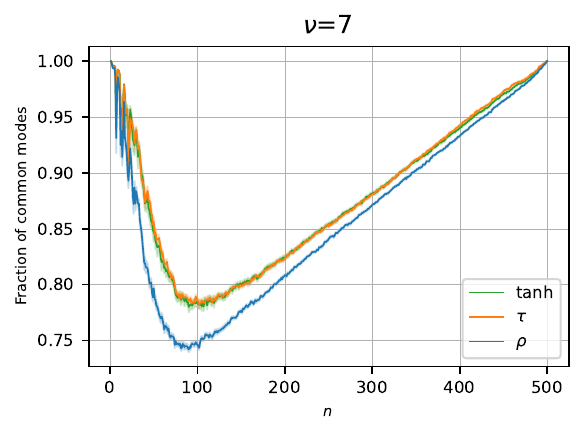}
  \end{minipage}
  \vspace{1em}
  \begin{minipage}{0.35\linewidth}
    \centering
    \includegraphics[width=\linewidth]{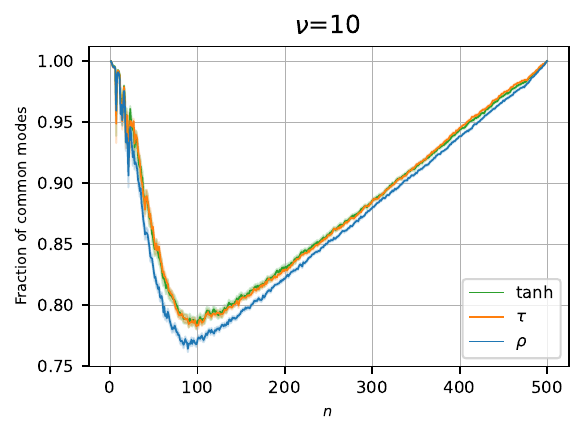} 
  \end{minipage}
  \begin{minipage}{0.35\linewidth}
    \centering
    \includegraphics[width=\linewidth]{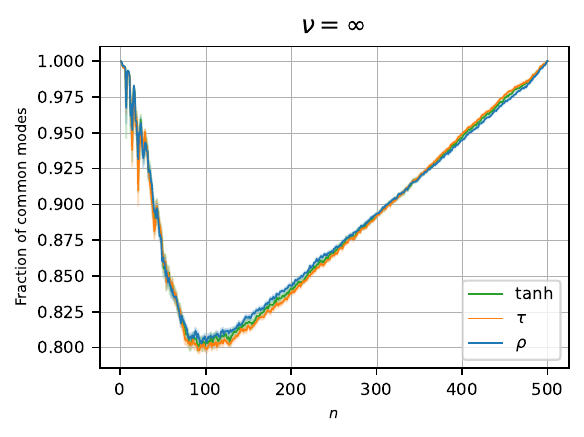} 
  \end{minipage}
  \caption{Fraction of common modes averaged over $50$ realizations for $q = 0.1$.} 
  \label{fig:focm_0.1}
\end{figure}

In this section, we conduct a more detailed analysis of the FCM between the true and estimated eigenvectors for multivariate Student data by varying the degrees of freedom $\nu$ and the parameter $q$. Our objective is to determine whether there exists a combination of parameters $(\nu, q)$ for which the Pearson FCM outperforms that of GCMs. Specifically, we anticipate that for relatively small values of $q$ (i.e., many data points) and large values of $\nu$ (approaching a Gaussian regime), the Pearson FCM will perform better, as the Pearson correlation matrix is particularly well-suited for Gaussian data. We aim to identify this potential phase transition. As for Fig.\ \ref{fig:focm}, we use the correlation matrix $\mathbf{C}$ of the 500 American stocks as the underlying correlation matrix for the synthetic data. We vary the degrees of freedom $\nu \in \{3, 7, 10, +\infty\}$ and sample sizes such that $q \in [0.1, 0.5]$. The results we present focus on the large modes, as the behavior is similar for the smaller modes. Specifically, we display results for $q=0.5$ (Fig.\ \ref{fig:focm_0.5}) and $q=0.1$ (Fig.\ \ref{fig:focm_0.1}). Notably, GCMs consistently perform as well as, or better than, Pearson in these cases. Even for multivariate Gaussian data (i.e., $\nu = +\infty$), the performance remains comparable. We have numerically verified that this behavior persists across different matrix sizes, with the number of entries varying between 500 and 50. Remarkably, even for $50 \times 50$ correlation matrices with 5000 observations (i.e., $q=0.01$) and Gaussian data, the FCMs of GCMs and Pearson remain close (GCMs have higher variance).

\FloatBarrier
\section{Average Out-of-Sample Risk} \label{app:tables_risk}

In the tables below, the top three methods, based on average performance and standard deviation, are highlighted in gold, silver, and bronze. When the ranking between several methods is unclear, e.g., lower average but higher variance, the methods are assigned the same color.

\begin{table}[htbp]
  \caption{\justifying Annualized average volatility $\left \langle \sigma_{\mathrm{out}} \right \rangle$ (in \textperthousand \, or \%) of the
    different strategies for $T=1000$, i.e., $q=0.5$. Standard deviations are given in bracket.}
  \label{tab:q_05}

  \renewcommand{\arraystretch}{1.2} 
\hspace{-38pt}
  \begin{minipage}{0.45\textwidth}
    \centering
    \begin{tabular}{|c|c|c|c|}
      \hline
      \multicolumn{4}{|c|}{\textbf{Omniscient}}                                                                                                                                                         \\

      \hline
      $\left \langle \sigma_{\mathrm{out}} \right \rangle $& America (\textperthousand)& APAC (\%)& Europe (\%)\\
      \hline
      Kendall                                                       & \cellcolor[RGB]{255, 215, 0}13.74 (4.45)   & \cellcolor[RGB]{255, 215, 0}31.32 (13.22)& \cellcolor[RGB]{255, 215, 0}26.66 (11.60)\\

      Kendall Clipped                                               & \cellcolor[RGB]{205, 127, 50}14.10 (4.78)  & \cellcolor[RGB]{205, 127, 50}31.75 (13.54)& \cellcolor[RGB]{205, 127, 50}26.84 (12.04)\\

      Kendall ICVC                                                  & \cellcolor[RGB]{192, 192, 192}13.82 (4.54) & \cellcolor[RGB]{192, 192, 192}31.36 (13.36)& \cellcolor[RGB]{255, 215, 0}26.56 (11.81)\\

      \hline

      RIE                                                           & 14.49 (5.28)                               & 32.44 (13.84)& 28.45 (12.37)\\

      RIE $\Gamma$                                                  & 14.47 (5.40)                               & 32.10 (13.94)& 27.30 (12.21)\\

      RIE + Id                                                      & 14.49 (5.28)                               & 32.43 (13.84)& 28.47 (12.39)\\

      Clipped                                                       & 14.79 (5.68)                               & 32.32 (14.19)& 27.28 (12.57)\\

      ICVC                                                          & 14.48 (5.38)                               & 32.17 (13.99)& 27.56 (12.24)\\

      \hline

      \multicolumn{4}{|c|}{\textbf{Random Long-Short}}                                                                                                                                                  \\
      \hline
      $\left \langle \sigma_{\mathrm{out}} \right \rangle$          & America (\textperthousand)& APAC (\%)& Europe (\%)\\

      \hline

      Kendall                                                       & \cellcolor[RGB]{205, 127, 50}3.18 (0.18)& 8.01 (0.28)                               & 7.62 (0.44)                               \\


      Kendall Clipped                                               & \cellcolor[RGB]{192, 192, 192}3.17 (0.18)& \cellcolor[RGB]{192, 192, 192}7.82 (0.28) & \cellcolor[RGB]{192, 192, 192}7.35 (0.45) \\

      Kendall ICVC                                                  & \cellcolor[RGB]{255, 215, 0}3.09 (0.17)& \cellcolor[RGB]{255, 215, 0}7.70 (0.27)   & \cellcolor[RGB]{255, 215, 0}7.29 (0.48)   \\

      \hline

      RIE                                                           & 3.58 (0.22)& 8.65 (0.54)                               & 8.84 (0.58)                               \\

      RIE $\Gamma$                                                  & 3.25 (0.18)& 7.91 (0.29)                               & 7.85 (0.45)                               \\

      RIE + Id                                                      & 3.58 (0.22)& 8.62 (0.52)                               & 8.87 (0.59)                               \\

      Clipped                                                       & 3.21 (0.18)& \cellcolor[RGB]{205, 127, 50}7.88 (0.26)  & \cellcolor[RGB]{205, 127, 50}7.42 (0.46)  \\

      ICVC                                                          & 3.26 (0.18)& 8.00 (0.35)                               & 7.81 (0.43)                               \\

      \hline
    \end{tabular}
  \end{minipage}
  \hspace{30pt}
  \begin{minipage}{0.45\textwidth}
    \centering
    \begin{tabular}{|c|c|c|c|}
      \hline
      \multicolumn{4}{|c|}{\textbf{Minimum Variance}}                                                                                                                                                   \\
      \hline

      $\left \langle \sigma_{\mathrm{out}} \right \rangle$          & America (\textperthousand)& APAC (\%)& Europe (\%)\\

      \hline

      Kendall                                                       & \cellcolor[RGB]{205, 127, 50}33.43 (8.62)& \cellcolor[RGB]{255, 215, 0}89.92 (13.48)& \cellcolor[RGB]{192, 192, 192}86.26 (16.83)\\


      Kendall Clipped                                               & \cellcolor[RGB]{192, 192, 192}33.30 (8.80)& \cellcolor[RGB]{192, 192, 192}90.67 (13.75)& \cellcolor[RGB]{205, 127, 50}87.52 (16.84)\\

      Kendall ICVC                                                  & \cellcolor[RGB]{255, 215, 0}33.24 (8.68)& \cellcolor[RGB]{205, 127, 50}91.49 (13.54)& \cellcolor[RGB]{255, 215, 0}86.31 (16.67)\\

      \hline

      RIE                                                           & 39.24 (11.42)& 96.18 (17.18)& 94.16 (20.77)\\

      RIE $\Gamma$                                                  & 38.92 (11.44)& 95.13 (17.08)& 93.62 (21.21)\\

      RIE + Id                                                      & 39.23 (11.42)& 96.13 (17.18)& 94.18 (20.76)\\

      Clipped                                                       & 37.84 (11.06)& 95.60 (16.83)& 92.96 (20.85)\\

      ICVC                                                          & 38.90 (11.43)& 95.20 (17.18)& 93.47 (21.25)\\

      \hline

      \multicolumn{4}{|c|}{\textbf{Mean Reversion}}                                                                                                                                                     \\
      \hline

      $\left \langle \sigma_{\mathrm{out}} \right \rangle $& America (\textperthousand)& APAC (\%)& Europe (\%)\\

      \hline

      Kendall                                                       & \cellcolor[RGB]{255, 215, 0}12.03 (3.96)& \cellcolor[RGB]{255, 215, 0}27.22 (6.29)& \cellcolor[RGB]{255, 215, 0}29.52 (8.16)\\


      Kendall Clipped                                               & \cellcolor[RGB]{205, 127, 50}12.36 (4.06)& \cellcolor[RGB]{205, 127, 50}27.99 (6.40)& \cellcolor[RGB]{205, 127, 50}29.77 (8.18)\\

      Kendall ICVC                                                  & \cellcolor[RGB]{192, 192, 192}12.07 (3.98)& \cellcolor[RGB]{192, 192, 192}27.39 (6.32)& \cellcolor[RGB]{192, 192, 192}29.52 (8.18)\\

      \hline

      RIE                                                           & 13.50 (0.89)& 28.72 (6.77)& 33.60 (9.10)\\

      RIE $\Gamma$                                                  & 13.37 (4.86)& 28.29 (6.70)& 31.59 (8.67)\\

      RIE + Id                                                      & 13.50 (4.80)& 28.71 (6.77)& 33.64 (9.12)\\

      Clipped                                                       & 13.57 (4.98)& 28.68 (6.82)& 30.93 (8.89)\\

      ICVC                                                          & 13.40 (0.88)& 28.41 (6.74)& 32.07 (8.87)\\

      \hline
    \end{tabular}
  \end{minipage}
\end{table}

\begin{table}[htbp]
  \caption{\justifying Annualized average volatility $\left \langle \sigma_{\mathrm{out}} \right \rangle$ (in \textperthousand \, or \%) of the different strategies for $T = 500$, i.e., $q=1$. Standard deviations are given in
    bracket.}
  \label{tab:q_1}

  \renewcommand{\arraystretch}{1.2} 

\hspace{-46pt}
  \begin{minipage}{0.45\textwidth}
    \centering
    \begin{tabular}{|c|c|c|c|}
      \hline
      \multicolumn{4}{|c|}{\textbf{Omniscient}}                                                                                                                                                         \\
      \hline
      $\left \langle \sigma_{\mathrm{out}} \right \rangle $& America (\textperthousand)& APAC (\%)& Europe (\%)\\
      \hline
      Kendall                                                       & \cellcolor[RGB]{255, 215, 0}14.37 (5.49)& \cellcolor[RGB]{255, 215, 0}31.32 (11.74)& \cellcolor[RGB]{255, 215, 0}27.80 (12.02)\\
      Kendall Clipped                                               & \cellcolor[RGB]{205, 127, 50}14.71 (5.74)& \cellcolor[RGB]{205, 127, 50}32.05 (12.52)& \cellcolor[RGB]{205, 127, 50}28.01 (12.36)\\
      Kendall ICVC                                                  & \cellcolor[RGB]{192, 192, 192}14.48 (5.60)& \cellcolor[RGB]{192, 192, 192}31.45 (12.06)& \cellcolor[RGB]{255, 215, 0}27.70 (12.32)\\
      \hline
      RIE                                                           & 21.47 (7.64)& 52.90 (13.71)& 62.84 (19.99)\\

      RIE $\Gamma$                                                  & 15.40 (6.92)& 32.06 (12.83)& 28.30 (12.91)\\
      RIE + Id                                                      & 17.13 (6.60)& 36.92 (12.46)& 38.31 (11.92)\\
      Clipped                                                       & 15.82 (7.06)& 32.44 (13.01)& 28.60 (12.89)\\
      ICVC                                                          & 15.33 (6.84)& 32.06 (12.83)& 28.27 (12.86)\\
      \hline
      \multicolumn{4}{|c|}{\textbf{Random Long-Short}}                                                                                                                                                  \\
      \hline
      $\left \langle \sigma_{\mathrm{out}} \right \rangle$          & America (\textperthousand)                                  & APAC (\%)& Europe (\%)\\
      \hline
      Kendall                                                       & 3.38 (0.19)& 8.49 (0.30)                               & 8.08 (0.46)                               \\
      Kendall Clipped                                               & \cellcolor[RGB]{192, 192, 192}3.21 (0.17)& \cellcolor[RGB]{205, 127, 50}7.93 (0.28)& \cellcolor[RGB]{192, 192, 192}7.44 (0.45) \\
      Kendall ICVC                                                  & \cellcolor[RGB]{255, 215, 0}3.17 (0.17)& \cellcolor[RGB]{255, 215, 0}7.82 (0.27)   & \cellcolor[RGB]{255, 215, 0}7.39 (0.45)   \\
      \hline
      RIE                                                           & 10.27 (0.73)& 33.28 (2.38)& 46.75 (4.70)\\

      RIE $\Gamma$                                                  & \cellcolor[RGB]{205, 127, 50}3.25 (0.19)& \cellcolor[RGB]{205, 127, 50}7.94 (0.27)& 7.58 (0.46)\\
      RIE + Id                                                      & 6.36 (0.40)& 15.82 (0.84)                              & 20.6 (1.52)                               \\
      Clipped                                                       & 3.29 (0.18)& 7.97 (0.26)                               & \cellcolor[RGB]{205, 127, 50}7.48 (0.45)  \\
      ICVC                                                          & 3.27 (0.19)& \cellcolor[RGB]{192, 192, 192}7.92 (0.26) & 7.61 (0.45)                               \\
      \hline
    \end{tabular}
  \end{minipage}
  \hspace{30pt}
  \begin{minipage}{0.45\textwidth}
    \centering
    \begin{tabular}{|c|c|c|c|}
      \hline
      \multicolumn{4}{|c|}{\textbf{Minimum Variance}}                                                                                                                                                   \\
      \hline
      $\left \langle \sigma_{\mathrm{out}} \right \rangle$          & America (\textperthousand)& APAC (\%)& Europe (\%)\\
      \hline
      Kendall                                                       & \cellcolor[RGB]{205, 127, 50}34.27 (9.50)& \cellcolor[RGB]{255, 215, 0}89.50 (14.68)& \cellcolor[RGB]{205, 127, 50}85.62 (18.75)\\
      Kendall Clipped                                               & \cellcolor[RGB]{192, 192, 192}34.09 (9.84)& 92.48 (16.25)& \cellcolor[RGB]{192, 192, 192}85.43 (18.59)\\
      Kendall ICVC                                                  & \cellcolor[RGB]{255, 215, 0}33.83 (9.58)& \cellcolor[RGB]{192, 192, 192}89.70 (15.47)& \cellcolor[RGB]{255, 215, 0}85.05 (18.90)\\
      \hline
      RIE                                                           & 49.39 (12.96)& 230.65 (27.39)& 222.88 (35.15)\\

      RIE $\Gamma$                                                  & 38.92 (12.82)& 91.81 (20.31)& 93.56 (25.63)\\
      RIE + Id                                                      & 42.49 (13.29)& 109.57 (19.14)& 136.78 (24.92)\\
      Clipped                                                       & 38.72 (11.85)& 93.47 (20.22)& 94.09 (25.24)\\
      ICVC                                                          & 38.99 (12.89)& 91.91 (20.29)& 93.74 (25.50)\\
      \hline
      \multicolumn{4}{|c|}{\textbf{Mean Reversion}}                                                                                                                                                     \\
      \hline
      $\left \langle \sigma_{\mathrm{out}} \right \rangle $& America (\textperthousand)& APAC (\%)& Europe (\%)\\
      \hline
      Kendall                                                       & \cellcolor[RGB]{255, 215, 0}12.49 (4.13)& \cellcolor[RGB]{255, 215, 0}27.92 (6.61)& \cellcolor[RGB]{255, 215, 0}29.53 (7.92)\\
      Kendall Clipped                                               & \cellcolor[RGB]{205, 127, 50}12.92 (4.35)& \cellcolor[RGB]{205, 127, 50}29.16 (7.05)& \cellcolor[RGB]{205, 127, 50}29.86 (8.31)\\
      Kendall ICVC                                                  & \cellcolor[RGB]{192, 192, 192}12.64 (4.23)& \cellcolor[RGB]{192, 192, 192}28.23 (6.82)& \cellcolor[RGB]{192, 192, 192}29.53 (8.21)\\
      \hline
      RIE                                                           & 19.82 (4.29)& 50.31 (7.50)& 66.51 (9.86)\\

      RIE $\Gamma$                                                  & 14.09 (5.21)& 29.02 (7.36)& 30.79 (8.98)\\
      RIE + Id                                                      & 15.58 (4.50)& 33.47 (6.87)& 40.69 (8.60)\\
      Clipped                                                       & 14.81 (5.46)& 29.57 (7.51)& 30.70 (9.10)\\
      ICVC                                                          & 13.99 (5.15)& \cellcolor[RGB]{205, 127, 50}29.02 (7.35)& 30.78 (8.92)\\
      \hline
    \end{tabular}
  \end{minipage}
\end{table}

\begin{table}[htpb]
  \centering
  \caption{\justifying Annualized average volatility $\left \langle \sigma_{\mathrm{out}} \right \rangle$ (in \textperthousand \, or \%) of the
    different strategies for $T=250$, i.e., $q=2$. Standard deviations are given in bracket.}
  \label{tab:q_2}

  \renewcommand{\arraystretch}{1.2} 
\hspace{-48pt}
  \begin{minipage}{0.45\textwidth}
    \centering
    \begin{tabular}{|c|c|c|c|}
      \hline
      \multicolumn{4}{|c|}{\textbf{Omniscient}}                                                                                                                                                                                    \\
      \hline
      $\left \langle \sigma_{\mathrm{out}} \right \rangle $& America (\textperthousand)& APAC (\%)& Europe (\%)\\
      \hline
      Kendall                                                       & \cellcolor[RGB]{255, 215, 0}14.39 (4.76)& \cellcolor[RGB]{205, 127, 50}33.00 (13.34)& \cellcolor[RGB]{255, 215, 0}28.61 (12.30)\\
      Kendall Clipped                                               & \cellcolor[RGB]{205, 127, 50}14.78 (5.37)& \cellcolor[RGB]{192, 192, 192}32.62 (13.25)& \cellcolor[RGB]{205, 127, 50}28.79 (13.17)\\
      Kendall ICVC                                                  & \cellcolor[RGB]{192, 192, 192}14.64 (5.23)& \cellcolor[RGB]{205, 127, 50}\cellcolor[RGB]{255, 215, 0}32.26 (13.04)& \cellcolor[RGB]{255, 215, 0}28.52 (13.01)\\
      \hline
      RIE                                                           & $\sim10^2$ ($\sim10^3$)& $\sim 10^2$ ($\sim10^2$)& $\sim10^2$ ($\sim10^2$)\\

      RIE $\Gamma$                                                  & 31.05 (7.69)& 57.33 (16.73)& 62.43 (18.61)\\
      RIE + Id                                                      & 15.43 (5.76)& 33.48 (13.20)& 30.06 (13.63)\\
      Clipped                                                       & 16.48 (7.05)& 33.37 (13.79)& 29.69 (14.25)\\
      ICVC                                                          & 15.72 (6.45)& 32.97 (13.86)& 29.40 (14.15)\\
      \hline
      \multicolumn{4}{|c|}{\textbf{Random Long-Short}}                                                                                                                                                                             \\
      \hline
      $\left \langle \sigma_{\mathrm{out}} \right \rangle$          & America (\textperthousand)& APAC (\%)& Europe (\%)\\
      \hline
      Kendall                                                       & 3.58 (0.17)& 10.57 (0.65)                                                         & 8.34 (0.47)                               \\
      Kendall Clipped                                               & \cellcolor[RGB]{192, 192, 192}3.25 (0.17)& \cellcolor[RGB]{192, 192, 192}8.07 (0.27)                            & \cellcolor[RGB]{192, 192, 192}7.50 (0.44) \\
      Kendall ICVC                                                  & \cellcolor[RGB]{255, 215, 0}3.23 (0.17)& \cellcolor[RGB]{255, 215, 0}7.99 (0.27)                              & \cellcolor[RGB]{255, 215, 0}7.46 (0.44)   \\
      \hline
      RIE                                                           & $\sim 10^2$ ($\sim 10^1$)& 77.54 (4.75)& $\sim 10^2$ ($\sim10^2$)\\

      RIE $\Gamma$                                                  & 18.29 (1.53)& 38.86 (2.81)& 42.58 (3.48)\\
      RIE + Id                                                      & 4.07 (0.24)& 9.87 (0.32)                                                          & 9.41 (0.49)                               \\
      Clipped                                                       & \cellcolor[RGB]{205, 127, 50}3.39 (0.18)& 8.21 (0.26)                                                          & \cellcolor[RGB]{205, 127, 50}7.58 (0.44)  \\
      ICVC                                                          & 3.36 (0.20)& \cellcolor[RGB]{205, 127, 50}8.08 (0.27)                             & \cellcolor[RGB]{205, 127, 50}7.59 (0.43)\\
      \hline
    \end{tabular}
  \end{minipage}
  \hspace{40pt}
  \begin{minipage}{0.45\textwidth}
    \centering
    \begin{tabular}{|c|c|c|c|}
      \hline
      \multicolumn{4}{|c|}{\textbf{Minimum Variance}}                                                                                                                                                                                 \\
      \hline
      $\left \langle \sigma_{\mathrm{out}} \right \rangle$          & America (\textperthousand)& APAC (\%)& Europe (\%)\\
      \hline
      Kendall                                                       & \cellcolor[RGB]{205, 127, 50}35.46 (10.42)& \cellcolor[RGB]{205, 127, 50}93.15 (16.04)& \cellcolor[RGB]{205, 127, 50}84.75 (19.99)\\
      Kendall Clipped                                               & \cellcolor[RGB]{192, 192, 192}35.03 (10.98)& \cellcolor[RGB]{192, 192, 192}91.09 (17.13)& \cellcolor[RGB]{255, 215, 0}84.53 (19.36)\\
      Kendall ICVC                                                  & \cellcolor[RGB]{255, 215, 0}34.82 (10.88)& \cellcolor[RGB]{255, 215, 0}88.98 (17.00)& \cellcolor[RGB]{192, 192, 192}\cellcolor[RGB]{192, 192, 192}84.82 (19.86)\\
      \hline
      RIE                                                           & 202.88 (23.6)& 347.94 (43.23)& 428.72 (50.4)\\

      RIE $\Gamma$                                                  & 69.11 (11.90)& 215.28 (23.57)& 199.30 (34.30)\\
      RIE + Id                                                      & 41.52 (14.69)& 98.81 (22.22)& 100.94 (28.86)\\
      Clipped                                                       & 41.27 (14.31)& 94.73 (21.89)& 94.55 (27.78)\\
      ICVC                                                          & 40.92 (14.51)& 93.72 (22.39)& 95.67 (28.71)\\
      \hline
      \multicolumn{4}{|c|}{\textbf{Mean Reversion}}                                                                                                                                                                                   \\
      \hline
      $\left \langle \sigma_{\mathrm{out}} \right \rangle $& America (\textperthousand)& APAC (\%)& Europe (\%)\\
      \hline
      Kendall                                                       & \cellcolor[RGB]{255, 215, 0}13.09 (4.40)& \cellcolor[RGB]{192, 192, 192}30.35 (6.24)& \cellcolor[RGB]{255, 215, 0}29.74 (8.34)\\
      Kendall Clipped                                               & \cellcolor[RGB]{205, 127, 50}13.63 (4.82)& \cellcolor[RGB]{192, 192, 192}30.23 (7.06)& \cellcolor[RGB]{205, 127, 50}30.47 (9.02)\\
      Kendall ICVC                                                  & \cellcolor[RGB]{192, 192, 192}13.47 (4.76)& \cellcolor[RGB]{255, 215, 0}29.61 (6.86)& \cellcolor[RGB]{192, 192, 192}30.16 (8.89)\\
      \hline
      RIE                                                           & $\sim10^3$ ($\sim 10^4$)& $\sim 10^2$ ($\sim 10^2$)& $\sim10^3$ ($\sim10^4$)\\

      RIE $\Gamma$                                                  & 30.68 (4.48)& 56.30 (10.88)& 63.99 (10.37)\\
      RIE + Id                                                      & 14.85 (5.51)& 30.61 (7.08)& 32.15 (9.73)\\
      Clipped                                                       & 16.30 (6.59)& 31.33 (7.72)& 31.82 (10.32)\\
      ICVC                                                          & 15.25 (6.10)& 30.61 (7.67)& 31.58 (10.17)\\
      \hline
    \end{tabular}
  \end{minipage}
\end{table}

\FloatBarrier

\begin{figure}
  \centering
  \includegraphics[width=0.65\linewidth]{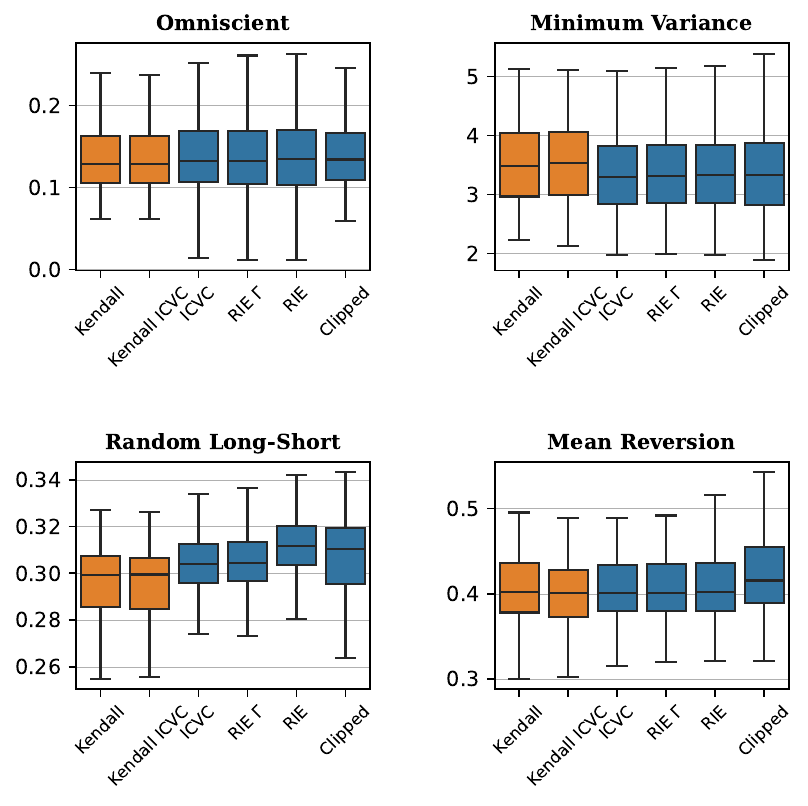}
  \caption{Annualized volatility $\left \langle \sigma_{\mathrm{out}} \right \rangle$ (in \%) of the different strategies for $q=0.2$ (American stocks).}
  \label{fig:ofs_risk_q_0.2}
\end{figure}

Finally, following the approach outlined in Appendix \ref{app:focm}, we investigate the performance of GCMs as the parameter \( q \) decreases, aiming to identify the threshold where their performance becomes comparable to the other methods. For American stocks, we find that performance remains better up to approximately \( q \approx 0.2 \) (see Fig.\ \ref{fig:ofs_risk_q_0.2}), while for European and APAC stocks, this threshold is closer to \( q \approx 0.4 \). As illustrated in Fig.\ \ref{fig:ofs_risk_q_0.2}, it is unclear whether Kendall-based methods outperform other strategies for American stocks at \( q=0.2 \), with the exception of the random long-short strategy. The following observation can be made: while omniscient and minimum variance strategies effectively model realized risk (with uncertainty only in the correlations), the random long-short strategy assesses the quality of estimated risk for a portfolio with uniform exposure to all modes. At \( q=0.2 \), both Kendall and Pearson methods exhibit similar accuracy in realized risk estimation, but Kendall shows lower risk for the uniform portfolio, highlighting its superior ability to estimate eigenvectors across the entire spectrum (this observation is further substantiated by Tables \ref{tab:q_05}, \ref{tab:q_1}, and \ref{tab:q_2} for the random long-short strategy, where Kendall ICVC consistently shows lower risk).

\end{document}